\documentclass[twocolumn,floatfix,superscriptaddress,showpacs,showkeys]{revtex4}
\usepackage{graphicx}
\usepackage{amsmath}
\usepackage{booktabs}
\usepackage{amssymb}
\usepackage{multirow}
\usepackage{makecell}
\usepackage{textcomp}
\usepackage{subfigure}
\usepackage{phonetic}
\usepackage{extarrows}
\usepackage{color}
\usepackage{float}
\usepackage[colorlinks,citecolor=magenta,linkcolor=magenta]{hyperref}
\begin{document}
\title{Quantum transport in a one-dimensional quasicrystal with mobility edges}
\author{Yan Xing}
\affiliation{School of Physics, Harbin Institute of Technology, Harbin, Heilongjiang 150001, China}
\author{Lu Qi}
\affiliation{School of Physics, Harbin Institute of Technology, Harbin, Heilongjiang 150001, China}
\author{Xuedong Zhao}
\affiliation{School of Physics, Harbin Institute of Technology, Harbin, Heilongjiang 150001, China}
\author{Zhe L\"{u}}
\email{lvzhe@hit.edu.cn}
\affiliation{School of Physics, Harbin Institute of Technology, Harbin, Heilongjiang 150001, China}
\author{Shutian Liu}
\affiliation{School of Physics, Harbin Institute of Technology, Harbin, Heilongjiang 150001, China}
\author{Shou Zhang}
\email{szhang@ybu.edu.cn}
\affiliation{Department of Physics, College of Science, Yanbian University, Yanji, Jilin 133002, China}
\author{Hong-Fu Wang}
\email{hfwang@ybu.edu.cn}
\affiliation{Department of Physics, College of Science, Yanbian University, Yanji, Jilin 133002, China}

\begin{abstract}
Quantum transport in a one-dimensional (1D) quasiperiodic lattice with mobility edges is explored. We first investigate the adiabatic pumping between left and right edge modes by resorting to two edge-bulk-edge channels and demonstrate that the success or failure of the adiabatic pumping depends on whether the corresponding bulk subchannel undergoes a localization-delocalization transition. Compared with the paradigmatic Aubry-Andr\'e (AA) model, the introduction of mobility edges triggers an opposite outcome for successful pumping in the two channels, showing a discrepancy of critical condition, and facilitates the robustness of the adiabatic pumping against quasidisorder. We also consider the transfer between excitations at both boundaries of the lattice and an anomalous phenomenon characterized by the enhanced quasidisorder contributing to the excitation transfer is found. Furthermore, there exists a parametric regime where a nonreciprocal effect emerges in the presence of mobility edges, which leads to a unidirectional transport for the excitation transfer and enables potential applications in the engineering of quantum diodes.
\pacs{05.60.Gg, 71.23.Ft, 72.20.Ee}
\keywords{quantum transport, quasicrystal, mobility edge}
\end{abstract}

\maketitle

\section{Introduction}\label{sec1}
Anderson localization~\cite{PhysRev.109.1492,RevModPhys.57.287,RevModPhys.80.1355}, a fundamental and ubiquitous phenomenon in nature, describes the absence of electron diffusion and the nonergodicity of electronic Bloch waves aroused by disorder, which originates from destructive quantum interference and uncovers the underlying mechanism behind metal-insulator transitions~\cite{Roati2008,PhysRevLett.103.013901}. According to scaling theory of localization~\cite{PhysRevLett.42.673}, while arbitrarily small random (uncorrelated) disorder in both one and two dimensions produces spatially exponential localization for all single-particle states, there exists a localization-delocalization transition in higher dimensions and an energy-dependent mobility edge distinguishing localized regime from delocalized regime appears at the phase boundary. However, this picture is altered dramatically in quasicrystals~\cite{PhysRevLett.53.2477,PhysRevLett.54.1730,PhysRevB.34.596,PhysRevB.34.617} when replacing the random disorder with a quasiperiodic potential. One of the most celebrated examples is characterized by the 1D AA model~\cite{Harper1955,AA1980,PhysRevB.14.2239}, which displays such a energy-independent critical behavior at a self-dual point. Moreover, rapid developments and remarkable advances in the manipulation of ultracold atoms loaded into incommensurate bichromatic optical lattices render a powerful tool to experimentally realize the AA model~\cite{Roati2008,Billy2008,Kondov66,Jendrzejewski2012,PhysRevLett.111.145303}. Interestingly, by incorporating further-range hopping~\cite{PhysRevA.80.021603,PhysRevLett.104.070601,PhysRevB.83.075105,PhysRevB.96.054202,PhysRevB.98.104201,PhysRevLett.123.025301,PhysRevB.100.174201} or appropriately designing on-site modulation~\cite{PhysRevLett.114.146601,PhysRevB.96.174207,PhysRevLett.125.196604}, some modified AA models can also sustain energy-dependent mobility edges. Recently, the experimental observations of mobility edges have also been reported in ultracold atom systems ~\cite{PhysRevLett.120.160404,PhysRevLett.122.170403,PhysRevLett.126.040603}.

On the other hand, quasicrystals exhibit topologically nontrivial phase~\cite{PhysRevLett.108.220401,PhysRevLett.109.116404,PhysRevLett.110.076403} that is attributed to higher-dimensional systems~\cite{PhysRevLett.109.106402}, for instance, the 1D AA model under an open boundary geometry supports gapless end modes spanning the bulk gaps, which are equivalent to the edge modes existing in the two-dimensional quantum Hall effect~\cite{PhysRevB.88.125118}. With the concept of topological protection in mind, these modes localized on the open boundary are immune to local perturbations and disorder, such as thermal fluctuations and defects, making them become a promising candidate in exploiting noise-resistant quantum information processing~\cite{Alicea2011,Lang2017} and fault-tolerant quantum computation~\cite{RevModPhys.80.1083,Stern1179}. More importantly, from a dynamical point of view, perhaps Thouless pumping~\cite{PhysRevB.27.6083,Nakajima2016,Lohse2016,PhysRevB.100.064302,PhysRevA.101.052323}, the quantization of the excitation transport, which occurs if the parameters of the system are adiabatically tuned in a cyclic manner, is the most visual reflection for the nontrivial topology of the system since the generated quantized excitation transport per cycle is in response to the topological invariants of the system. Furthermore, the topological pumping of boundary modes~\cite{PhysRevLett.109.106402,PhysRevB.91.014108,Zhao2017,PhysRevB.91.064201,PhysRevB.101.094307,PhysRevB.102.014305,PhysRevLett.125.224301,PhysRevLett.126.095501} can be another significant domain and has attracted intensive attention. It also depends on an adiabatic adjustment of the parameters of the system and dedicates an excitation transport across the bulk, which opens up avenues for robust quantum state transfer~\cite{Yao2013,Dlaska_2017,PhysRevA.98.012331,Longhi2019,PhysRevB.99.155150,PhysRevResearch.2.033475,PhysRevB.102.174312,PhysRevA.102.012606,PhysRevA.102.022404,PhysRevA.103.023504,PhysRevB.103.085129,PhysRevA.103.032402,PhysRevResearch.3.023037,cao2021}, resilient quantum gate preparation~\cite{PhysRevB.100.045414}, etc. 

Although the investigation on these basic principles and essential properties has got substantial progress in their respective realms, the interplay among them still remains less explored, e.g., the effect of Anderson localization and mobility edges on the adiabatic pumping between boundary modes is still not explicitly understood and clarified and how mobility edges further affect the excitation transport is not clearly demonstrated yet, and quasicrystals provide an ideal platform to address these problems. 

Motivated by the confusion, in this work, we explore the quantum transport in a 1D quasiperiodic lattice with mobility edges. Firstly, the adiabatic pumping between left and right edge modes is investigated by resorting to two edge-bulk-edge channels, which is always feasible before the Anderson localization of the corresponding bulk subchannel occurs, and the relevant cause for the failure of the adiabatic pumping in the localized regime is analyzed and discussed in detail. Compared with the AA model where successful pumping in both the channels can be achieved until the system undergoes a localization-delocalization transition, an opposite outcome for successful pumping in the two channels is disclosed with mobility edges being introduced, while one can survive only under a reduced critical condition, the other can persist beyond the constraint of the AA model, which, in a sense, indicates the presence of mobility edges facilitates the robustness of the adiabatic pumping against quasidisorder. Depending on the pumping process, the transfer between excitations at both boundaries of the lattice is also considered and found to be anomalous in the delocalized regime showing the enhanced quasidisorder raises fidelity of the excitation transfer. Moreover, there exists a parametric regime where a nonreciprocal transport manifested by the unidirectional excitation transfer is captured, which is induced by mobility edges and cannot emerge in the AA model. 

The remainder of this paper is organized as follows. In Sec.~\ref{sec2}, we introduce the model and present the existence of mobility edges for a sample system. In Sec.~\ref{sec3}, for both the delocalized and localized regimes, the adiabatic pumping between left and right edge modes in the AA model is first explored by resorting to two edge-bulk-edge channels, where we detailedly analyze and discuss what determines the success or failure of the adiabatic pumping in terms of the energy spectrum of the system. Subsequently, the adiabatic pumping in the sample system is shown and the effect of mobility edges on the adiabatic pumping in each channel is expounded. We also quantify the pumping outcome based on fidelity. In Sec.~\ref{sec4}, from the perspective of the application of the adiabatic pumping, we consider the transfer between excitations at both boundaries of the lattice, where some novel transport phenomena are found. A conclusion is given in Sec.~\ref{sec5}. 

\section{Model and mobility edges}\label{sec2}
\begin{figure}
\centering
\includegraphics[width=0.9\linewidth]{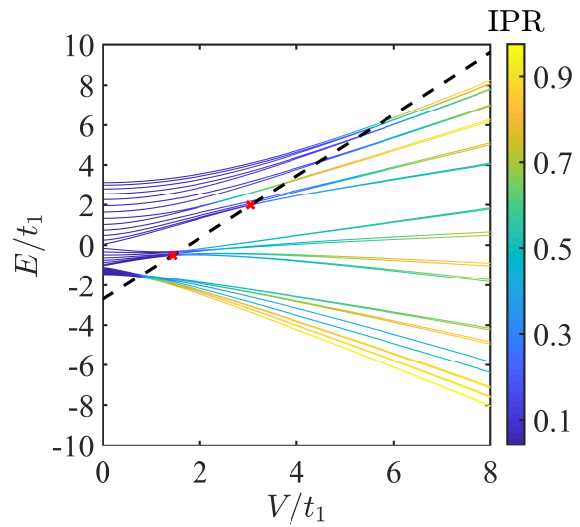}
\caption{(Color online) Energy spectrum of a sample system with $u=1$ versus modulation strength $V$ for $\varphi=0.99\pi$. The black (dashed) line delineates the analytical result of the mobility edge, which is validated according to IPR of each eigenmode, as indicated by the color bar. The two red crosses mark the threshold $V_{c}$ for the bulk subchannels of channels $A$ and $B$, respectively.}
\label{fig1}
\end{figure}

We consider a 1D shallow bichromatic optical lattice, where the ratio between the periods of the primary and secondary cosine potentials is incommensurate, so that the non-nearest-neighbor tight-binding framework is released and a lattice model subject to a short-range hopping can be attained~\cite{PhysRevA.80.021603,PhysRevLett.104.070601,PhysRevB.83.075105}. Specifically, we focus on a lattice model characterized by an exponentially decaying hopping and the Hamiltonian of the system reads 
\begin{eqnarray}
H=\sum_{m\neq l}te^{-u\left|m-l\right|}\left(a_{m}^{\dagger}a_{l}+a_{l}^{\dagger}a_{m}\right)+\sum_{m}w_{m}a_{m}^{\dagger}a_{m},
\label{e1}
\end{eqnarray}
here $a_{m}^{\dagger}$ ($a_{m}$) denotes the creation (annihilation) operator of a particle on site $m$. The exponentially decaying hopping is described by hopping amplitude $t$ and decaying coefficient $u>0$. The on-site potential takes the quasiperiodic cosine modulation $w_{m}=V\cos\left(2\pi\zeta m+\varphi\right)$ with parameters $V>0$, $\zeta^{-1}=\left(\sqrt{5}+1\right)/2$, and $\varphi\in\left[0,2\pi\right]$, respectively, being the modulation strength, period, and phase. The lattice model is just a simple short-range hopping counterpart of the AA model and the only inclusion of the hopping term $t_{1}=te^{-u}$, which is referred to as the energy unit throughout the paper, reduces the lattice model to the AA model. However, compared with the AA model that undergoes a localization-delocalization transition determined by a self-dual condition $V_{c}=2$, the lattice model satisfies a generalized self-dual symmetry and sustains energy-dependent mobility edges, which separate localized regime from delocalized regime at a critical energy $E_{c}$ governed by $E_{c}=V_{c}\cosh\left(u\right)-t$~\cite{PhysRevLett.104.070601}. Figure~\ref{fig1}, for example, depicts the energy spectrum of a sample system versus $V$ when $u=1$ and the black (dashed) line delineates the analytical result of the mobility edge. The presence of the mobility edge is validated according to inverse participation ratio (IPR) of each eigenmode, which is defined as
\begin{eqnarray}
\mathrm{IPR}=\frac{\sum_{m=1}^{N}|\psi_{m}|^{4}}{(\sum_{m=1}^{N}|\psi_{m}|^{2})^{2}},
\label{e2}
\end{eqnarray} 
where $\psi_{m}$ stands for the probability amplitude of the corresponding eigenmode at the $m$th site and $N$ is the total number of sites. One can readily confirm that $\mathrm{IPR}=1$ for the most localized state and $\mathrm{IPR}=1/N$ for the completely delocalized state. Visibly, the abrupt change of the IPR agrees well with the prediction of the mobility edge, which indicates that the localization-delocalization transition is dictated by the mobility edge indeed. 

\section{Robust adiabatic pumping facilitated by mobility edges}\label{sec3}
\begin{figure*}
\centering
\includegraphics[width=0.329\linewidth]{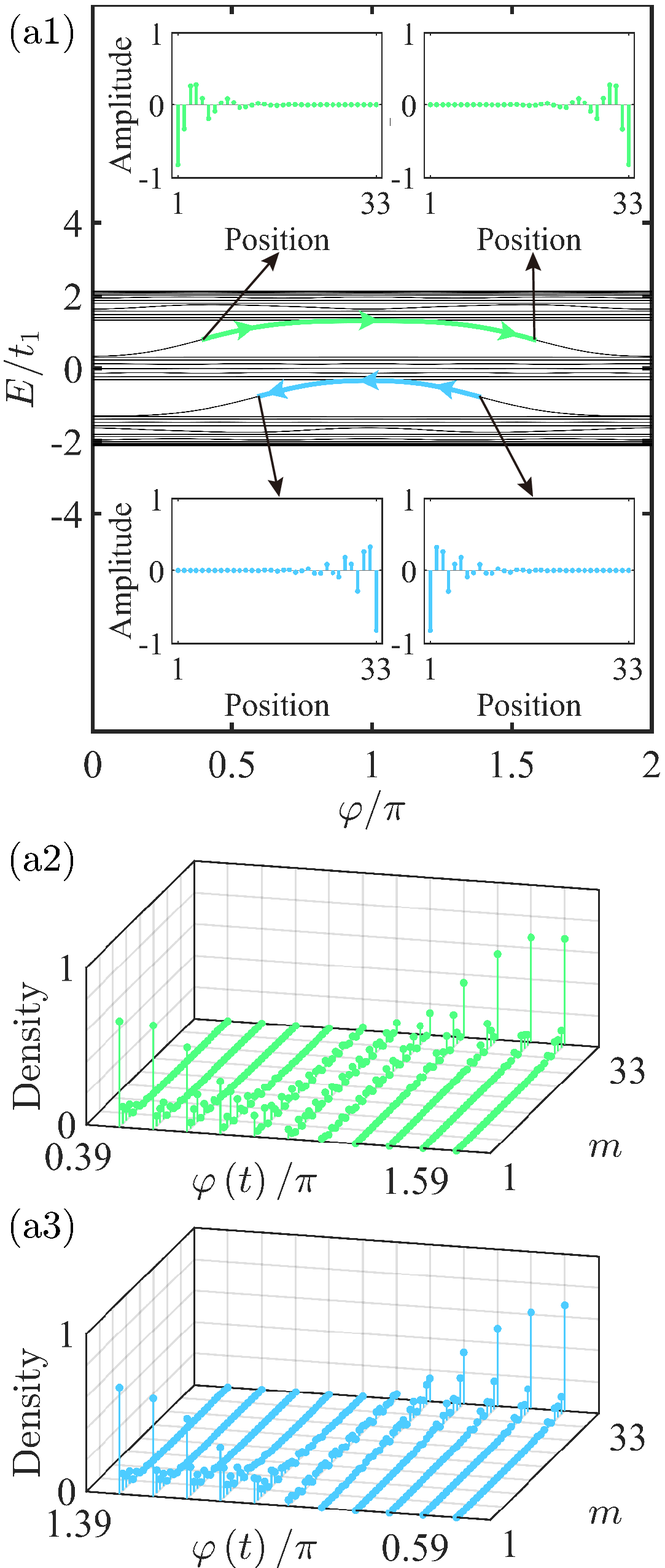}
\includegraphics[width=0.329\linewidth]{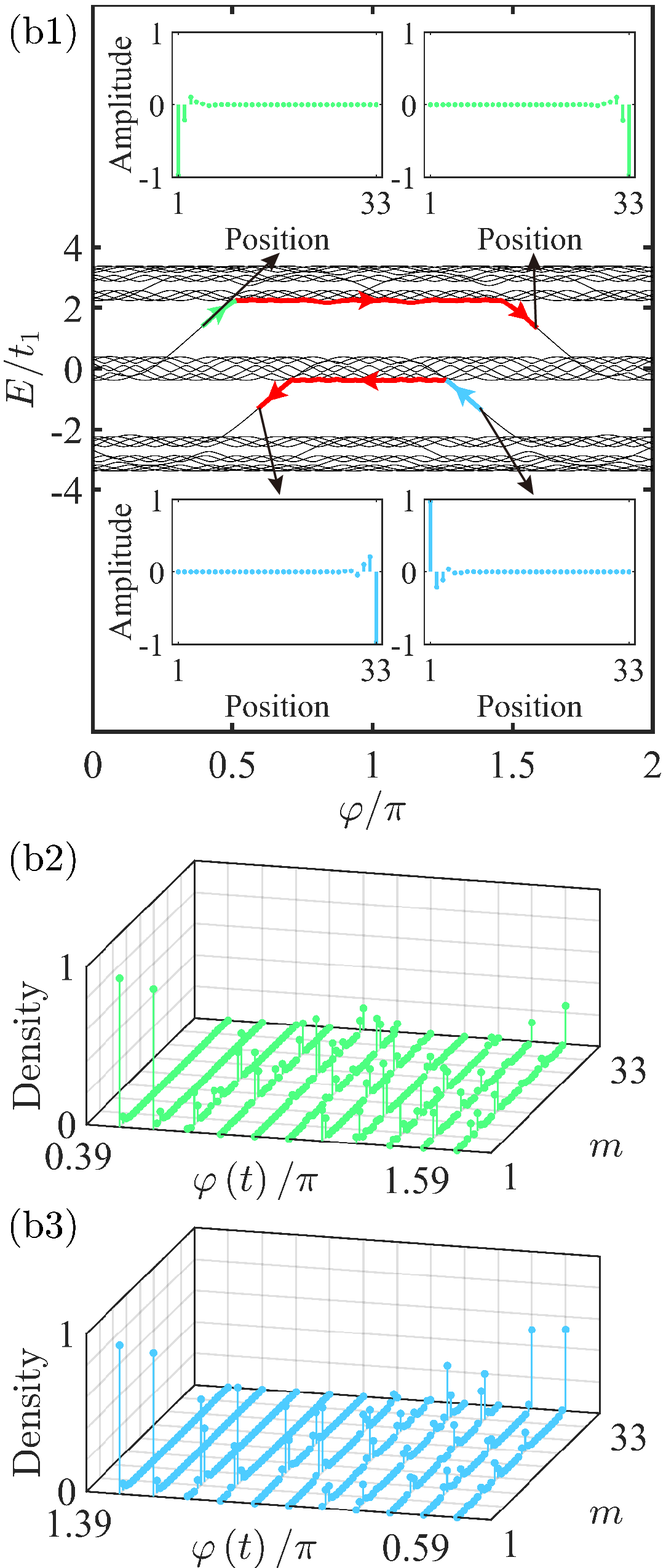}
\includegraphics[width=0.329\linewidth]{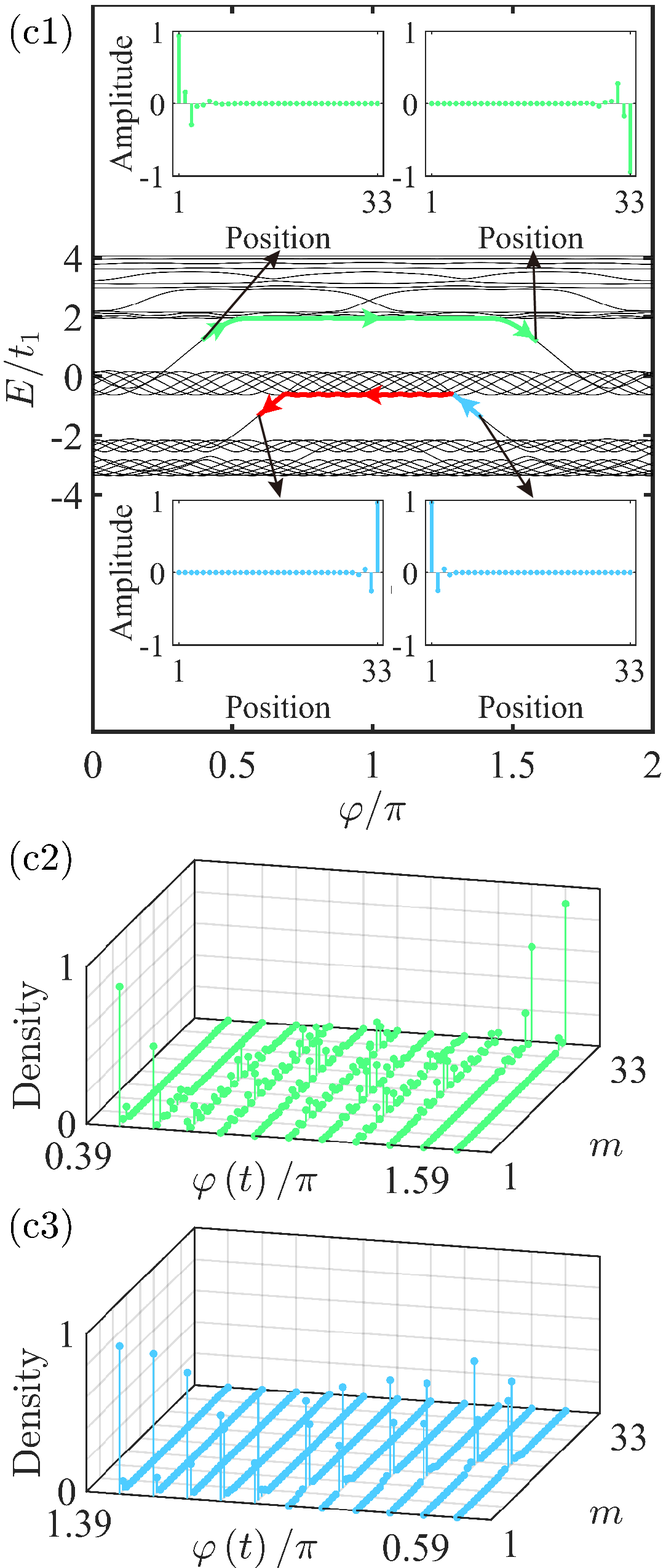}
\caption{(Color online) Energy spectrum of the AA model with the number of sites $N=33$ as a function of modulation phase $\varphi$ for (a1) $V=1$ and (b1) $V=3$. In (a1), the system is in delocalized regime. The green (upper) and blue (lower) lines label channel $A$ with $\varphi\in\left[0.39\pi, 1.59\pi\right]$ and channel $B$ with $\varphi\in\left[0.59\pi, 1.39\pi\right]$ for an adiabatic pumping between left and right edge modes, respectively. The arrows in each channel guide the corresponding pumping direction. In (b1), the system is in localized regime. Both bulk and right-edge subchannels break down in either channel, as indicated by the red (dark gray) line and arrow. (c1) Energy spectrum of the sample system as a function of modulation phase $\varphi$. Other parameters are the same as (b1). The channel $A$ remains intact but the channel $B$ is collapsed in this case. The insets in (a1)-(c1) display the initial left and goal right edge states used in the adiabatic evolution of each channel. The corresponding pumping process by resorting to (a2)-(c2) channel $A$ and (a3)-(c3) channel $B$. Here, $\Omega=10^{-5}$ in all the figures.}
\label{fig2}
\end{figure*}

\begin{figure}
\centering
\includegraphics[width=1.0\linewidth]{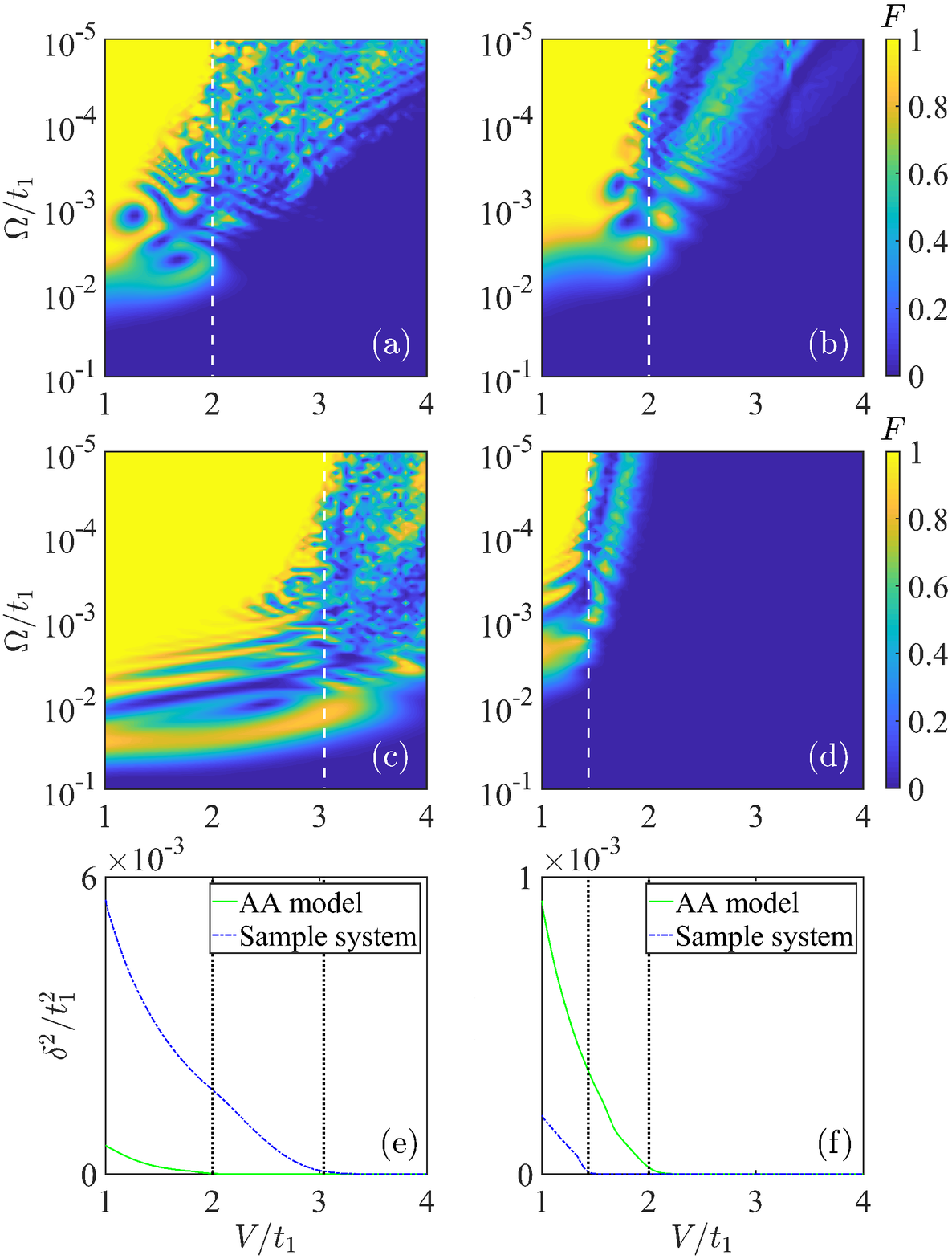}
\caption{(Color online) Fidelity $F$ of the adiabatic pumping between left and right edge modes versus modulation strength $V$ and ramping frequency $\Omega$ via (a) channel $A$ in the AA model, (b) channel $B$ in the AA model, (c) channel $A$ in the sample system, and (d) channel $B$ in the sample system. The square of the minimum gap $\delta$ between the corresponding bulk subchannel and its adjacent sublevel as a function of modulation strength $V$ for (e) channel $A$ and (f) channel $B$ of the AA model and sample system. The white (dashed) or black (dotted) lines in each figure delineate the threshold $V_{c}$ of the corresponding bulk subchannel extracted from Fig.~\ref{fig1} or $V=2$.}
\label{fig3}
\end{figure}

It is well known that the AA model supports topologically nontrivial edge modes that traverse the bulk gaps~\cite{PhysRevLett.109.106402,PhysRevB.88.125118}. Under open boundary condition (OBC), we present the energy spectrum of the AA model as a function of $\varphi$ with $V=1$ in Fig.~\ref{fig2}(a1). One can observe that there are two major gaps and gapless edge modes emerge within either gap. As displayed in insets of Fig.~\ref{fig2}(a1), the edge modes inside the upper (lower) gap are mainly concentrated on the left (right) boundary of the system if these modes cross from bottom to top, whereas the right (left) edge modes can be captured when the crossing direction becomes opposite. Furthermore, these edge modes become extended gradually as they merge into the bulk sublevels on both sides. Clearly, these left and right edge modes inside the upper (lower) gap are bridged by a bulk sublevel, which is at the bottom of these upper (middle) sublevels.

Due to the existence of the gap between this bulk sublevel and its adjacent sublevel, the system in the delocalized regime admits an adiabatic pumping between the left and right edge modes by engineering a relevant channel when $\varphi$ is scanned sufficiently slowly. As an example, if we intend to pump a left edge mode to the right boundary of the lattice, there are two feasible channels to implement. As shown in Fig.~\ref{fig2}(a1), one is labeled by the green (upper) line and called channel $A$, the other is labeled by the blue (lower) line and called channel $B$, and the arrows in each channel guide the corresponding pumping direction. We can divide either channel into three parts, namely, left-edge subchannel, bulk subchannel, and right-edge subchannel. Because the region of the bulk subchannel for either channel becomes wider with the increase of $V$, in order that there always exist the left- and right-edge subchannels for each channel in the investigated range of $V$, for the channel $A$ ($B$), e.g., $\varphi$ is slowly swept from $0.39\pi$ ($1.39\pi$) to $1.59\pi$ ($0.59\pi$) with $\varphi(t)=0.39\pi+\Omega t$ ($\varphi(t)=1.39\pi-\Omega t$), where $t$ represents the evolution time and $\Omega$ is the ramping frequency dominating the evolution velocity, and an adiabatic change in $\varphi$ has been experimentally realized in ultracold atom systems~\cite{Nakajima2016,Lohse2016}. We can obtain the pumping process by solving the Schr\"odinger equation numerically ($\hbar=1$),
\begin{eqnarray}
i\frac{d}{dt}|\Phi(t)\rangle=H(t)|\Phi(t)\rangle,
\label{e3}
\end{eqnarray}
here $\Phi(t)$ is spanned by basis $\{|m\rangle=a_{m}^{\dagger}|\mathbf{0}\rangle, 1\leqslant m\leqslant N\}$, i.e., $\Phi(t)=\sum^{N}_{m=1}\phi_{m}(t)|m\rangle$ with $\phi_{m}(t)$ being the probability amplitude on the $m$th site at time $t$. The corresponding pumping processes by resorting to the channels $A$ and $B$ are reported in Figs.~\ref{fig2}(a2) and \ref{fig2}(a3), respectively. Obviously, for both the channels, the whole pumping process undergoes three stages manifested by left-localized intensity $\rightarrow$ extended intensity $\rightarrow$ right-localized intensity, which means that the pumping entirely occupy each channel all the time and successively passes the three subchannels during the whole adiabatic evolution. Moreover, the left edge mode is pumped to the desired right edge mode successfully. 

The Anderson localization occurs once $V>2$ and all the bulk modes of the system become exponentially localized~\cite{AA1980}. The energy spectrum of the AA model as a function of $\varphi$ with $V=3$ under OBC is exhibited in Fig.~\ref{fig2}(b1). Different from the energy spectrum in the delocalized regime, except for these two major gaps, the gap between every two sublevels in the localized regime is squeezed and becomes exponentially suppressed at the multiple locations of $\varphi$, including the gap between each bulk subchannel and its adjacent sublevel, while these edge modes remain inside both the major gaps and are unaffected, which, in a sense, reflects the sensitivity of a bulk mode without topological protection and the immunity of a topologically protected edge mode to quasidisorder. Moreover, the localization of these edge modes is enhanced, as displayed in insets of Fig.~\ref{fig2}(b1). 

If the adiabatic pumping is executed again, we can find that the pumping for both the channels fails, as reported in Figs.~\ref{fig2}(b2) and~\ref{fig2}(b3). The underlying mechanism behind the failure of either pumping is attributed to the exponentially suppressed gap leading to the Landau-Zener tunneling between sublevels and the tunneling probability depends on the size of the avoided crossing and on the evolution velocity. Concretely, the narrower the avoided crossing or the faster the evolution velocity, the larger the tunneling probability. Therefore, the pumping completely enters into the bulk subchannel from the left-edge subchannel. Subsequently, when the pumping first encounters the avoided crossing at an exponentially suppressed gap between the bulk subchannel and its adjacent sublevel, depending on the magnitude of both the avoided crossing at the exponentially suppressed gap and $\Omega$, it could be subject to a partial or even whole tunneling from the bulk subchannel to its adjacent sublevel and further to the nonadiabatic excitation to the adjacent sublevel and thereby to the leak into the adjacent sublevel. The pumping will, hence, no longer occupy the bulk subchannel completely in later evolution. As evolution continues, whenever the pumping encounters the avoided crossing at an exponentially suppressed gap between sublevels, either a partial or a whole Landau-Zener tunneling between the two sublevels will occur. After undergoing a series of avoided crossings, there is a large possibility for the pumping to fail to return into the right-edge subchannel perfectly, as labeled by the red (dark gray) line and arrow for each channel in Fig.~\ref{fig2}(b1). Furthermore, an examination for these two pumping processes in Figs.~\ref{fig2}(b2) and~\ref{fig2}(b3) reveals the observation that the left-localized intensity transforms into several localized intensities at different sites of the lattice during evolution, which implies the occurrence of the Landau-Zener tunneling between sublevels and thus a simultaneous occupation of several exponentially localized bulk modes.   

It can be seen from the above analysis and discussion that the bulk subchannel is fragile and breaks down in the localized regime, which is responsible for the failure of either pumping. Consequently, the adiabatic pumping between the left and right edge modes survives until the occurrence of the Anderson localization, or in other words, the adiabatic pumping is not robust against quasidisorder with disorder strength $V>2$. In order to facilitate the robustness of the adiabatic pumping against quasidisorder, namely, achieving successful pumping under stronger quasidisorder, a crucial factor is to prevent the Landau-Zener tunneling from the bulk subchannel to its adjacent sublevel. Accordingly, a resolved gap between them should be guaranteed when $V>2$, which needs to raise the threshold $V_{c}$ for their localization-delocalization transition such that they can still be in the delocalized regime after $V>2$. On the other hand, with the introduction of the mobility edge, we can find in Fig.~\ref{fig1} that the thresholds of these upper (middle) sublevels are all greater (less) than $2$. More importantly, the bulk subchannel of the channel $A$ ($B$) is at the bottom of these upper (middle) sublevels, whose $V_{c}$ is approximately equal to $3.042$ ($1.434$), as marked by the red crosses. As a result, we will expect that successful pumping can always be achieved by resorting to the channel $A$ as long as $V<3.042$, whereas the adiabatic pumping in the channel $B$ survives only before $V>1.434$.    

In Fig.~\ref{fig2}(c1), we also plot the energy spectrum of the sample system as a function of $\varphi$ with $V=3$ under OBC to verify the effect of the mobility edge on both the bulk subchannels. One can observe that while the gaps within these middle sublevels all become exponentially suppressed and do not persist, the mobility edge maintains a resolved gap between every two upper sublevels, which demonstrates that the Anderson localization for these middle sublevels has already occurred but these upper sublevels have not undergone the localization-delocalization transition yet and are still in the delocalized regime. As a consequence, the bulk subchannel of the channel $A$ ($B$) remains intact (breaks down). Additionally, the corresponding pumping processes by resorting to the channels $A$ and $B$ are also reported in Figs.~\ref{fig2}(c2) and \ref{fig2}(c3), respectively. As expected, the adiabatic pumping in the channel $B$ fails, nevertheless, the left edge mode is successfully pumped to the desired right edge mode by resorting to the channel $A$ and successful pumping is achieved indeed when $V>2$, which contrasts sharply to the case of the AA model with the same $V$ shown in Fig.~\ref{fig2}(b2).   

For the purpose of elucidating the effect of the mobility edge on the adiabatic pumping more clearly, we introduce fidelity to quantitatively characterize the pumping outcome, which can be evaluated as follows:
\begin{eqnarray}
F=|\langle\Psi_{g}|\Psi_{f}\rangle|^{2},
\label{e4}
\end{eqnarray} 
where $|\Psi_{f}\rangle$ and $|\Psi_{g}\rangle$ are the normalized final state when the adiabatic evolution for an initial left edge state finishes and the normalized goal right edge state, respectively, and $0\leqslant F\leqslant1$. It is explicit that $F\rightarrow0$ indicates the failure of the pumping, on the contrary, the pumping is successful if $F\rightarrow1$. 

As a comparison, we first show the fidelities $F$ of the channels $A$ and $B$ versus $V$ and $\Omega$ for the AA model in Figs.~\ref{fig3}(a) and~\ref{fig3}(b), the fidelities $F$ of both the channels versus $V$ and $\Omega$ for the sample system are also displayed in Figs.~\ref{fig3}(c) and~\ref{fig3}(d). Furthermore, for the AA model and sample system, figure~\ref{fig3}(e) (\ref{fig3}(f)) exhibits the square of the minimum gap $\delta$ between the bulk subchannel of the channel $A$ ($B$) and its adjacent sublevel as a function of $V$. The white (dashed) or black (dotted) lines in each figure delineate $V_{c}$ of the corresponding bulk subchannel extracted from Fig.~\ref{fig1} or $V=2$. One can observe that for the AA model, if $\Omega$ is small enough (corresponding to a sufficiently adiabatic pumping process), successful pumping in both the channels can be achieved until the system undergoes the localization-delocalization transition, viz., when $V<2$. For the sample system, by resorting to the channel $A$, if $V<3.042$, i.e., before the corresponding bulk subchannel suffers the Anderson localization, successful pumping always persists provided that the evolution velocity is sufficiently slow, however, despite the adiabatic pumping in the channel $B$ also surviving until the occurrence of the Anderson localization of the corresponding bulk subchannel, we can realize it only before $V>1.434$. Moreover, as $V$ increases, we can find that the minimum $\Omega$ required for successful pumping dwindles, which roots in the decline of $\delta^{2}$ such that a longer temporal evolution is necessary. In the localized regime, due to the almost vanishing $\delta^{2}$, the pumping nearly fails without a very high fidelity no matter how small $\Omega$ becomes. In a word, the mobility edge play a pivotal role in improving the fragility of the bulk subchannel of the channel $A$ and facilitate the robustness of the adiabatic pumping against quasidisorder. 

Note that finite time nonadiabatic excitations always exist during evolution, therefore, it is not possible to have perfect pumping $F=1$. However, in the delocalized regime, all of these fidelities in the region of successful pumping are almost greater than $0.99$ and we can thus ignore these very few nonadiabatic excitations. 

It is worth mentioning that in the delocalized regime, as the size of the system increases, the gap between each bulk subchannel and its adjacent sublevel gradually shrinks but still persists unlike the exponentially suppressed gap in the localized regime, which leads to that the fidelity of the adiabatic pumping will depend on the length of the lattice. For achieving successful pumping in a longer lattice, despite the extended bulk modes of the system, a slower variation of $\varphi$ is needed to avoid the Landau-Zener tunneling from the bulk subchannel to its adjacent sublevel during evolution. 

\section{Anomalous excitation transfer and nonreciprocal transport}\label{sec4}
\begin{figure}
\centering
\includegraphics[width=0.49\linewidth]{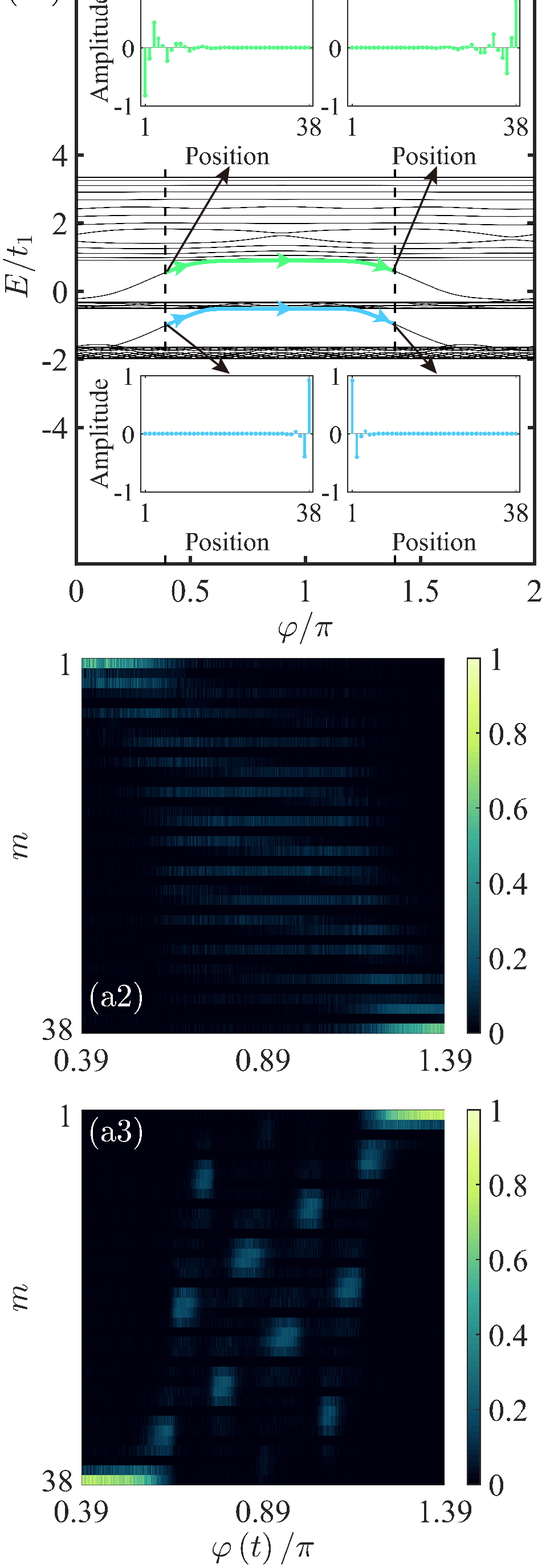}
\includegraphics[width=0.49\linewidth]{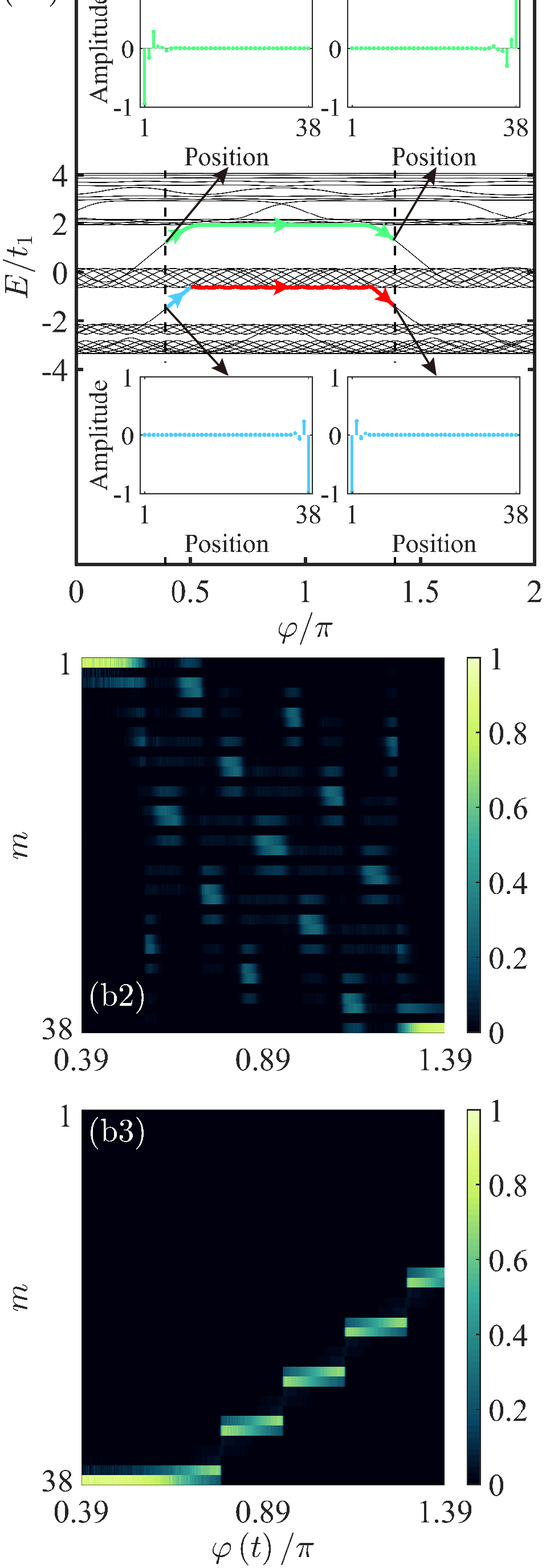}
\caption{(Color online) Energy spectrum of the sample system with the number of sites $N=38$ as a function of modulation phase $\varphi$ for (a1) $V=1.4$ and (b1) $V=3$. For the two channels, the adiabatic evolution is in the same direction and within the same region $\varphi\in\left[0.39\pi, 1.39\pi\right]$, as shown between the black (dashed) lines. The insets in (a1) and (b1) display the left and right edge modes at $\varphi=0.39\pi$ and $1.39\pi$. In (a1), both the channels remain intact, which results in a bidirectional transfer between excitations at both boundaries of the lattice, as reported in (a2) and (a3). In (b1), only the channel $A$ remains intact and the channel $B$ is collapsed, which brings about a nonreciprocal transport manifested by the unidirectional excitation transfer, as reported in (b2) and (b3). The color bar indicates the density on each site during evolution and $\Omega=10^{-5}$ in all the evolution processes.}
\label{fig4}
\end{figure} 

\begin{figure}
\centering
\includegraphics[width=1.0\linewidth]{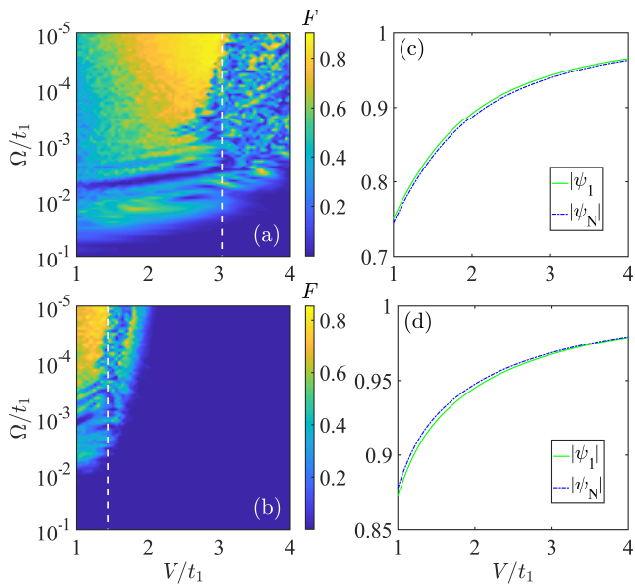}
\caption{(Color online) Fidelity $F$ of the excitation transfer in the sample system versus modulation strength $V$ and ramping frequency $\Omega$ via (a) channel $A$ and (b) channel $B$. The white (dashed) line in either figure delineates the threshold $V_{c}$ of the corresponding bulk subchannel, which is almost equal to that extracted from Fig.~\ref{fig1}. (c) In the channel $A$, $|\psi_{1}|$ of the left edge mode at $\varphi=0.39\pi$ and $|\psi_{N}|$ of the right edge mode at $\varphi=1.39\pi$ as a function of modulation strength $V$. (d) In the channel $B$, $|\psi_{N}|$ of the right edge mode at $\varphi=0.39\pi$ and $|\psi_{1}|$ of the left edge mode at $\varphi=1.39\pi$ as a function of modulation strength $V$.}
\label{fig5}
\end{figure}

The adiabatic pumping between the left and right edge modes could be potentially exploited in the transfer between excitations at both boundaries of the lattice, namely, quantum state transfer. To this end, we consider the initial state injected into the leftmost (rightmost) site for the channel $A$ ($B$). Because each edge mode is mainly concentrated on either side of the sample system, both the initial states mostly contain a superposition of a corresponding edge mode and the excitation transfer purely depends on the pumping process between the left and right edge modes. As exhibited in Fig.~\ref{fig4}(a1), for both the channels, the adiabatic evolution is in the same direction by slowly ramping $\varphi$ from $0.39\pi$ to $1.39\pi$ with $\varphi=0.39\pi+\Omega t$. The corresponding evolution processes by resorting to the channels $A$ and $B$ are reported in Figs.~\ref{fig4}(a2) and \ref{fig4}(a3), respectively. It is evident that an excitation is transferred to the other side of the sample system whether it is initially injected from the leftmost or rightmost sites, but the transfer outcome is imperfect, in other words, the transferred excitation is not entirely localized on the opposite boundary of the lattice. 

Similarly, we quantify the transfer outcome by employing fidelity, equation~\ref{e4}, where the goal state $|\Psi_{g}\rangle$ is chosen as the state injected into the rightmost (leftmost) site for the channel $A$ ($B$). Figures~\ref{fig5}(a) and~\ref{fig5}(b) exhibit the fidelities $F$ of the channels $A$ and $B$ versus $V$ and $\Omega$ for the sample system, respectively. One can observe that for both the channels, the occurrence of the Anderson localization also has an adverse effect on the excitation transfer caused by the collapse of the adiabatic pumping, which is feasible until the corresponding bulk subchannel undergoes the localization-delocalization transition. In the delocalized regime, we find that with the increase of $V$ and the decrease of $\Omega$, fidelity rises, which, in a sense, implies that the enhanced quasidisorder is conducive to the excitation transfer. This is unique for quasidisorder since increasing stochastic disorder strength will gradually eliminate the relevant gap and is generally not beneficial to quantum state transfer. The reason behind this phenomenon ascribes to the fact that the increasing quasidisorder strength $V$ enhances the localization of each edge mode on the corresponding boundary of the lattice, as displayed in Fig.~\ref{fig5}(c) (\ref{fig5}(d)) for $|\psi_{1}|$ ($|\psi_{N}|$) of the left (right) edge mode at $\varphi=0.39\pi$ and $|\psi_{N}|$ ($|\psi_{1}|$) of the right (left) edge mode at $\varphi=1.39\pi$, so that the weight of the left (right) edge mode in the eigenmode superposition of the initial state for the channel $A$ ($B$) is elevated. In consequence, as $V$ increases, although the relevant gap also gradually diminishes, when $\Omega$ is small enough, the pumping process between the left and right edge modes is more and more dominant in the whole adiabatic evolution, which contributes to receive a final state that is more localized on the other boundary of the lattice and further to a higher fidelity and thus to the anomalous excitation transfer. 

There exists another interesting phenomenon in the scope of $1.434<V<3.042$, viz., before the Anderson localization for the bulk subchannel of the channel $A$ occurs and after the bulk subchannel of the channel $B$ undergoes the localization-delocalization transition. It turns out that if we initially inject an excitation into the leftmost site, it can be adiabatically transfered to the opposite side of the sample system by scanning $\varphi$ sufficiently slowly except that the transferred excitation outputs incompletely from the rightmost site, nevertheless, when an excitation is initially injected into the rightmost site and $\varphi$ is slowly swept again within the same range, it is possible for the excitation not to output at all from the leftmost site, which arises from the presence of the mobility edge triggering an opposite outcome of the adiabatic pumping in this scope for the channels $A$ and $B$ and can be treated as a nonreciprocal effect, as shown in Fig.~\ref{fig4}(b1). The nonreciprocal effect will become more pronounced with $V$ approaching $3.042$, as reported in Figs.~\ref{fig4}(b2) and~\ref{fig4}(b3) with $V=3$, and cannot emerge in the AA model. Based on the nonreciprocal transport across the bulk induced by the mobility edge, it is feasible to engineer a quantum diode by viewing both the leftmost and rightmost sites as two ports. 

\section{Conclusions}\label{sec5}
In conclusion, we have investigated the quantum transport in a 1D quasicrystal with a specific short-range hopping, i.e., the adiabatic pumping between left and right edge modes and the transfer between excitations at both boundaries of the lattice. The quasiperiodic lattice is described by an exponentially decaying hopping and sustains energy-dependent mobility edges. The adiabatic pumping can be implemented by resorting to two channels and is always feasible before the occurrence of the Anderson localization of the corresponding bulk subchannel, which results from the exponentially suppressed gap in the localized regime leading to the Landau-Zener tunneling between sublevels. Therefore, for both the channels, successful pumping in the AA model can be achieved until the system undergoes the localization-delocalization transition, viz., when $V<2$. Compared with the AA model, the introduction of mobility edges increases (decreases) the threshold of the transition for the bulk subchannel of the channel $A$ ($B$), which brings about a different outcome for the adiabatic pumping in the two channels. In other words, successful pumping in the channel $A$ persists even after $V>2$ but in the channel $B$ survives only with a reduced critical condition. If we interpret the incommensurate on-site potential as a highly correlated disorder, overall, it turns out that the presence of mobility edges actually facilitates the robustness of the adiabatic pumping against quasidisorder. Furthermore, depending on the pumping process, we have shown that the transfer between excitations at both boundaries of the lattice is anomalous in the delocalized regime, which is characterized by the enhanced quasidisorder elevating fidelity of the excitation transfer. Moreover, there exists a parametric regime where a nonreciprocal effect emerges. The nonreciprocal effect results in a unidirectional transport for the excitation transfer, which is absent in the AA model. Based on the nonreciprocal transport, a promising application could be in the exploitation and preparation of quantum diodes.

Although the present work focuses on adiabatic evolution, we also hope that more interest can be stimulated for the exploration of the quantum transport in quasicrystals based on active controls, such as, shortcuts to adiabaticity.

\begin{center}
{\bf{ACKNOWLEDGMENTS}}
\end{center}
This work was supported by the National Natural Science Foundation of China under Grants No. 61822114, No. 12074330, No. 62071412, No. 11874132, and No. 61575055.

\end{document}